\documentclass{article}

\usepackage{amsmath,graphicx,mlspconf,soul,algorithm, algorithmic}
\usepackage{IEEEtrantools}
\usepackage[hidelinks]{hyperref}
\usepackage[capitalize]{cleveref}
\usepackage[inline]{enumitem}
\usepackage{{booktabs}}
\usepackage[font=normalsize,skip=4pt]{caption}
\usepackage{tikz}
\usetikzlibrary{shapes,arrows,positioning}

\newcommand\blfootnote[1]{%
  \begingroup
  \renewcommand\thefootnote{}\footnote{#1}%
  \addtocounter{footnote}{-1}%
  \endgroup
}

\copyrightnotice{979-8-3503-2411-2/23/\$31.00 {\copyright}2023 IEEE}

\toappear{2023 IEEE International Workshop on Machine Learning for Signal Processing, Sept.\ 17--20, 2023, Rome, Italy}

\title{Dynamic nsNet2: Efficient Deep Noise Suppression with Early Exiting}
 \name{
    \begin{tabular}{c c c c c c c c c}
\multicolumn{4}{c}{Riccardo Miccini$^{{\star}{\dagger}}$} & \multicolumn{5}{c}{Alaa Zniber$^{\mathsection}$} \\
\multicolumn{2}{c}{Clément Laroche$^{\star}$} & \multicolumn{2}{c}{Tobias Piechowiak$^{\star}$} & \multicolumn{2}{c}{Martin Schoeberl$^{\dagger}$} & \multicolumn{3}{c}{Luca Pezzarossa$^{\dagger}$} \\
\multicolumn{3}{c}{Ouassim Karrakchou$^{\mathsection}$} & \multicolumn{3}{c}{Jens Sparsø$^{\dagger}$} &  \multicolumn{3}{c}{Mounir Ghogho$^{\mathsection}$} \\
    \end{tabular}
 }
 \address{
    $^{\star}$ GN Audio
    \qquad $^{\dagger}$ Technical University of Denmark
    \qquad $^{\mathsection}$ International University of Rabat
 }

\begin{document}

\maketitle

\begin{abstract}
Although deep learning has made strides in the field of deep noise suppression, leveraging deep architectures on resource-constrained devices still proved challenging. 
Therefore, we present an early-exiting model based on nsNet2 that provides several levels of accuracy and resource savings by halting computations at different stages.
Moreover, we adapt the original architecture by splitting the information flow to take into account the injected dynamism. We show the trade-offs between performance and computational complexity based on established metrics. 

\end{abstract}

\blfootnote{This research has received funding from the European Union’s Horizon research and innovation programme under grant agreement No 101070374.}

\bstctlcite{IEEEexample:BSTcontrol}

\vspace{-0.1cm}
\section{Introduction}
\label{sec:intro}

In recent years, audio products such as earbuds, headsets, and hearing aids have become increasingly popular and are driving the demand for real-time solutions capable of improving speech quality. 
Noise suppression techniques based on deep learning --- also referred to as Deep Noise Suppression (DNS) --- are superseding conventional speech enhancement techniques based on digital signal processing. This is thanks to their ability to effectively reduce non-stationary noise and deal with diverse background sounds. Several DNS models, in particular those capable of real-time causal inference, rely on Recurrent Neural Networks (RNNs) \cite{braun_data_2020, tan_convolutional_2018, braun_towards_2021}.
Such approaches usually involve feeding a frequency-domain signal --- in the form of Short-time Fourier Transform (STFT) or Mel spectrograms --- one frame at a time, into an encoding layer and one or more recurrent layers such as Gated Recurrent Units (GRU), thereby providing temporal sequence modelling abilities. This internal representation is then decoded to form a time-varying filter, which is applied to the noisy input to obtain a clean speech estimate.

Unfortunately, such architectures are not particularly suited for resource-constrained devices due to the large amount of computation required by the recurrent units. On that note, several efforts have been put in trying to optimise and deploy recurrent models on embedded hardware \cite{rusci_accelerating_2022, fedorov_tinylstms_2020} with techniques such as quantization. Nevertheless, one fundamental aspect remains unaddressed, i.e., the model requires the same amount of computation irrespective of energy/power requirements, initial speech quality of the input signal, and desired output speech quality.

To address this issue, Dynamic Neural Networks (DyNNs), also known as conditional neural networks, are one path to undergo \cite{han_dynamic_2021}. DyNNs can adaptively adjust their parameters as well as their computational graph, based on the input they receive. This dynamism grants the network more flexibility and efficiency in handling various resource budgets, real-time requirements, and device capacities while maintaining a good performance trade-off. Amongst the most promising techniques for DyNNs that appear suitable for addressing limited hardware resources, we find \textit{early exiting} \cite{scardapane_why_2020, laskaridis_adaptive_2021, tan_empowering_2021}. Early exiting was introduced in the context of image classification \cite{huang_multi-scale_2022, kaya_shallow-deep_2019}. It appends the architecture with decision blocks (e.g., internal classifiers) that decide when to halt the computations based on the output of each exit stage (e.g., image class probability). This paradigm helps avoid using the full-fledged network, but various challenging aspects emerge from these new types of architectures. For instance, the absence of feature sharing between internal decision blocks can lead to computational waste \cite{zhou_bert_2020, wolczyk_zero_2021}. Another challenge lies in the optimal placement of the decision blocks in the architecture based on either performance gain or loss minimisation \cite{panda_energy-efficient_2017, baccarelli_optimized_2020}. Moreover, convergence is also subject to instability due to the  accumulation of gradients coming from the different decision blocks, which can be mitigated by gradient re-scaling \cite{li_improved_2019} or a mixture of training strategies \cite{brock_freezeout_2017}.

In this work, we introduce a novel application of DyNNs to noise suppression. We adapt nsNet2 \cite{braun_data_2020}, a popular architecture for noise suppression, into a model capable of early exiting given user-predefined constraints. We then demonstrate that the resulting architecture achieves a monotonic increase in its denoising capabilities with each consecutive exit stage. Our architecture allows the user to choose the denoising/computational cost trade-off that best suits their needs. Due to the additional challenges associated with unintrusively modelling speech quality in real-time, automatic exiting is beyond the scope of this work. In this paper:
\begin{itemize}
\vspace{-0.1cm}\item
  We convert nsNet2 into an early-exiting model and explore several architectural adaptations aimed at maintaining its denoising capabilities at each stage;
\vspace{-0.1cm}\item
  We investigate the impact of different early exit training strategies (layer-wise or joint) on the denoising performance of the models;
\vspace{-0.1cm}\item
  We evaluate the speech quality and computational efficiency of our models for each exit stage and show that our architectural adaptations decrease computational cost without degrading baseline performances.
\end{itemize}


\section{Deep noise suppression}
\label{sec:dns}

We assume that the observed signal $X(k,n)$ --- defined in the STFT-domain where $k$ and $n$ are the frequency and time frame indices, respectively --- is modelled as an additive mixture of the desired speech $S(k,n)$ and interfering noise $N(k,n)$, expressed as:
\begin{equation} \label{eq0}
X(k,n) = S(k,n) + N(k,n)
\end{equation}

To obtain the clean speech estimate $\widehat{S}(k,n)$, we compute a real-valued suppression gain spectral mask $\widehat{M}$ such that:
\begin{equation} \label{eq1}
\widehat{S}(k,n) = X(k,n) \cdot \widehat{M}(k,n)
\end{equation}

This can be formulated as a supervised learning task, where a mask $\widehat{M}(k,n)$ is learned to minimise the distance between $S$ and $\widehat{S}$. 

\subsection{Dynamic Architecture}
\label{ssec:arch}

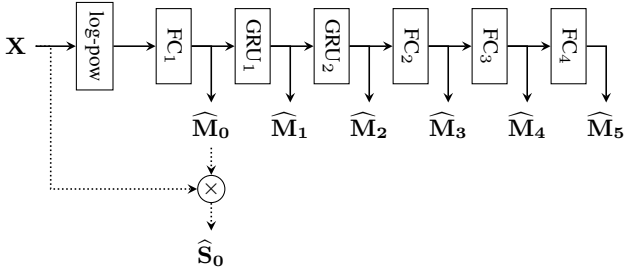
\begin{figure}[t]
\begin{minipage}[b]{0.95\linewidth}
  \centering
  \centerline{\begin{tikzpicture}[node distance=1.05cm]

\tikzstyle{inout} = [text centered, font=\small]
\tikzstyle{nn} = [rectangle, minimum width=1cm, minimum height=0.4cm, text centered, draw=black, font=\footnotesize]
\tikzstyle{line} = [semithick,->,>=stealth]

\node(in) [inout] {$\mathbf{X}$};
\node(preproc) [nn, right of=in, rotate=-90] {$\text{log-pow}$};
\node(FC1) [nn, right of=preproc, rotate=-90] {FC$_1$};
\node(GRU1) [nn, right of=FC1, rotate=-90] {GRU$_1$};
\node(GRU2) [nn, right of=GRU1, rotate=-90] {GRU$_2$};
\node(FC2) [nn, right of=GRU2, rotate=-90] {FC$_2$};
\node(FC3) [nn, right of=FC2, rotate=-90] {FC$_3$};
\node(FC4) [nn, right of=FC3, rotate=-90] {FC$_4$};

\node(M0hat) [inout, below of=FC1, xshift=0.5cm] {$\mathbf{\widehat{M}_0}$};
\node(M1hat) [inout, below of=GRU1, xshift=0.5cm] {$\mathbf{\widehat{M}_1}$};
\node(M2hat) [inout, below of=GRU2, xshift=0.5cm] {$\mathbf{\widehat{M}_2}$};
\node(M3hat) [inout, below of=FC2, xshift=0.5cm] {$\mathbf{\widehat{M}_3}$};
\node(M4hat) [inout, below of=FC3, xshift=0.5cm] {$\mathbf{\widehat{M}_4}$};
\node(M5hat) [inout, below of=FC4, xshift=0.5cm] {$\mathbf{\widehat{M}_5}$};

\node(prod) [circle, draw=black, font=\footnotesize, inner sep=0.02cm, below of=M0hat, yshift=0.2cm] {$\times$};
\node(S0hat) [inout, below of=prod, yshift=0.2cm] {$\mathbf{\widehat{S}_0}$};

\draw [line] (in) -- (preproc);
\draw [line] (preproc) -- (FC1);
\draw [line] (FC1) -- (GRU1);
\draw [line] (GRU1) -- (GRU2);
\draw [line] (GRU2) -- (FC2);
\draw [line] (FC2) -- (FC3);
\draw [line] (FC3) -- (FC4);

\draw [line] (FC1.north) -| (M0hat.north);
\draw [line] (GRU1.north) -| (M1hat.north);
\draw [line] (GRU2.north) -| (M2hat.north);
\draw [line] (FC2.north) -| (M3hat.north);
\draw [line] (FC3.north) -| (M4hat.north);
\draw [line] (FC4.north) -| (M5hat.north);

\draw [line, densely dotted] (M0hat.south) -- (prod.north);
\draw [line, densely dotted] (in.east) -| ++(0.2cm,0cm) |- (prod.west);
\draw [line, densely dotted] (prod.south) -- (S0hat.north);

\end{tikzpicture}}
\end{minipage}
  \caption{nsNet2 architecture with exit stages (dotted lines show an example of full inference path)}
  \label{fig:earlyexiting}
\end{figure}

As our baseline, we use nsNet2 \cite{braun_data_2020}, an established DNS model with the following architecture comprising recurrent and fully-connected (FC) layers: FC-GRU-GRU-FC-FC-FC. Each fully-connected layer is followed by a ReLU activation, except for the last one, which features a sigmoid nonlinearity. The model operates on real-valued log-power spectrograms, computed as $\log({|X|^2 + \varepsilon})$, for a small $\varepsilon$.

We introduce exit stages after each layer, as shown in \cref{fig:earlyexiting}. Each exit outputs a mask $\widehat{M}_i$, which is a subset of that layer's activations with size given by the number of frequency bins, to be applied to the noisy input (chosen, in our case, as the first 257 features).
Since the suppression gains are bound between 0 and 1, we use the sigmoid function to clamp our FC activations --- i.e., the outputs of layers FC$_1$, FC$_2$, FC$_3$ --- similarly to how the baseline model implements its last activation (i.e., the output of layer FC$_4$). 
For the GRU layer activations, which are inherently bound between $-1$ and $1$ due to the final $\tanh$ activation, we employ the simple scaling function $0.5 \cdot (1 + \text{GRU}_i(x))$.
Note that these extra steps are only performed when there is a need to extract the mask at the early stage, otherwise, we use the activations mentioned earlier.
This results in an architecture able to recover the signal with up to 6 different denoising abilities. 

Given the significant differences between the masks and the model's internal representation, forcing our model to derive the former at each layer may degrade its performance.
We address this by reducing the number of intermediate exit stages to promote the emergence of richer internal feature representations in non-exiting layers. Thus, we introduced a version of our dynamic model with only the 4 exits marked with $\mathbf{\widehat{M}_0}$, $\mathbf{\widehat{M}_1}$, $\mathbf{\widehat{M}_3}$, $\mathbf{\widehat{M}_5}$ in \cref{fig:earlyexiting}. We decided to remove later-occurring exits because, during our experimental observations, they were more prone to performance degradation. 


\subsection{Split Layers}
\label{ssec:splitlayers}

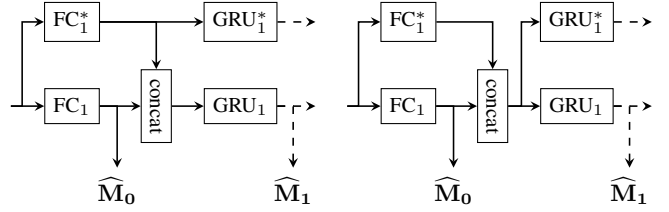
\begin{figure}[t]
\begin{minipage}[b]{0.48\linewidth}
  \centering
  \centerline{\begin{tikzpicture}[node distance=1.12cm]

\tikzstyle{nn} = [rectangle, minimum width=0.5cm, minimum height=0.4cm, text centered, draw=black, font=\footnotesize]
\tikzstyle{cat} = [rectangle, minimum width=0.5cm, minimum height=0.4cm, text centered, draw=black, font=\footnotesize]
\tikzstyle{exit} = [rectangle, minimum width=0.5cm, minimum height=0.4cm, text centered, font=\small]
\tikzstyle{line} = [semithick,->,>=stealth]

\node(in) [minimum width=0.6cm] {};
\node(FC1) [nn, right of=in] {FC$_1$};
\node(FCstar1) [nn, above of=FC1] {FC$^*_1$};
\node(concat) [cat, right of=FC1, rotate=-90] {concat};
\node(GRU1) [nn, right of=concat] {GRU$_1$};
\node(GRUstar1) [nn, above of=GRU1] {GRU$^*_1$};
\node(M0hat) [exit, below of=FC1, xshift=0.6cm] {$\mathbf{\widehat{M}_0}$};
\node(M1hat) [exit, below of=GRU1, xshift=0.7cm] {$\mathbf{\widehat{M}_1}$};
\node(out) [right of=GRU1] {};
\node(outstar) [right of=GRUstar1] {};

\draw [line] (in.east) -- (FC1);
\draw [line] (in.east) -| ++(0.15cm,1cm) |- (FCstar1.west);
\draw [line] (FC1.east) -| (M0hat.north);
\draw [line] (FC1) -- (concat);
\draw [line] (FCstar1) -| (concat);
\draw [line] (FCstar1) -- (GRUstar1);
\draw [line] (concat) -- (GRU1);
\draw [line, dashed] (GRUstar1) -- (outstar);
\draw [line, dashed] (GRU1) -- (out);
\draw [line, dashed] (GRU1.east) -| (M1hat.north);

\end{tikzpicture}}
  \centerline{(a) Regular split layer}
\end{minipage}
\hfill
\begin{minipage}[b]{0.48\linewidth}
  \centering
  \centerline{\begin{tikzpicture}[node distance=1.12cm]

\tikzstyle{nn} = [rectangle, minimum width=0.5cm, minimum height=0.4cm, text centered, draw=black, font=\footnotesize]
\tikzstyle{cat} = [rectangle, minimum width=0.5cm, minimum height=0.4cm, text centered, draw=black, font=\footnotesize]
\tikzstyle{exit} = [rectangle, minimum width=0.5cm, minimum height=0.4cm, text centered, font=\small]
\tikzstyle{line} = [semithick,->,>=stealth]

\node(in) [minimum width=0.6cm] {};
\node(FC1) [nn, right of=in] {FC$_1$};
\node(FCstar1) [nn, above of=FC1] {FC$^*_1$};
\node(concat) [cat, right of=FC1, rotate=-90] {concat};
\node(GRU1) [nn, right of=concat] {GRU$_1$};
\node(GRUstar1) [nn, above of=GRU1] {GRU$^*_1$};
\node(M0hat) [exit, below of=FC1, xshift=0.6cm] {$\mathbf{\widehat{M}_0}$};
\node(M1hat) [exit, below of=GRU1, xshift=0.7cm] {$\mathbf{\widehat{M}_1}$};
\node(out) [right of=GRU1] {};
\node(outstar) [right of=GRUstar1] {};

\draw [line] (in.east) -- (FC1);
\draw [line] (in.east) -| ++(0.15cm,1cm) |- (FCstar1.west);
\draw [line] (FC1.east) -| (M0hat.north);
\draw [line] (FC1) -- (concat);
\draw [line] (FCstar1) -| (concat);
\draw [line] (concat.north) -- ++(0.175cm,0cm) |- (GRUstar1.west);
\draw [line] (concat) -- (GRU1);
\draw [line, dashed] (GRUstar1) -- (outstar);
\draw [line, dashed] (GRU1) -- (out);
\draw [line, dashed] (GRU1.east) -| (M1hat.north);

\end{tikzpicture}}
  \centerline{(b) With concatenated inputs}
\end{minipage}

  \caption{Different styles of split layer adaptations}
  \label{fig:splitlayers}
\end{figure}

As mentioned earlier, when dealing with early exiting, a degradation in performance in the deepest exit stages is often observed, when comparing the model against its non-dynamic analogous. This is due to the emergence of \textit{task-specialized} features that prevent useful information to flow further \cite{han_dynamic_2021}. Therefore, we attempt to mitigate the issue by introducing additional data paths in the form of duplicate layers. These act as ancillary feature extractors, tasked with deriving an increasingly refined internal representation that proves useful for the downstream layers. This configuration also avoids subsetting the layers' activations to yield a mask.

As shown in \cref{fig:splitlayers}, two alternative split-layer topologies have been formulated. In both cases, the layers $\Phi_i$ generate mask estimates, while layers $\Phi_i^*$ propagate features. The main difference between the two variants is whether a given $\Phi_i^*$ layer receives only the output from $\Phi_{i-1}^*$ or the concatenated output from both previous layers, i.e., $concat(\Phi_{i-1}^{ }, \Phi_{i-1}^*)$. The former variant assumes that previous masks do not contain features that are useful for the model's internal representation. Conversely, the latter lifts this assumption at the expense of computational complexity, which is increased as a result of the larger input size for $\Phi_{i}^*$ layers.

\section{Training}
\label{sec:training}

Our model is trained on the loss function shown in \cref{eq2} that was initially proposed in \cite{wilson_exploring_2018} and subsequently adapted in \cite{braun_data_2020}. The loss comprises two terms: the first term computes the mean-squared error between the clean and estimated complex spectra, whereas the second term corresponds to the mean-squared error of the magnitude spectra. The two terms are weighted by a factor of $\alpha$ and $(1-\alpha)$, respectively, with $\alpha=0.3$. The spectra are power-law compressed with $c = 0.3$:

\begin{equation}
\label{eq2}
\begin{aligned}
\mathcal{L}\left(S,\widehat{S}\right) = {} & \alpha \sum_{k, n} \left| |S|^c e^{j \angle S} - |\widehat{S}|^c e^{j \angle \widehat{S}} \right|^2 + \\ 
& (1-\alpha) \sum_{k, n} \left| |S|^c- |\widehat{S}|^c\right|^2
\end{aligned}
\end{equation}

To avoid the impact of large signals dominating the loss and creating unbalanced training in a batch of several samples, we normalise $S$ and $\widehat{S}$ by the standard deviation of the target signal $\sigma_{S}$ before computing the loss, as per \cite{braun_data_2020}.

In general, training early-exit models can fall into two categories \cite{scardapane_why_2020}: \textit{layer-wise training} and \textit{joint training}. Since each has its advantages and drawbacks, we have adopted both strategies to challenge them against each other.

\subsection{Layer-wise Training}
\label{ssec:layerwise}

Layer-wise training is a straightforward way to train early exiting models with or without pre-trained backbones (i.e., non-dynamic architectures). The idea is to train the first sub-model, from input $X$ and target $S$ to the first exit stage that outputs an estimate $\widehat{S}_0 = X \odot \widehat{M}_0$. Once the training reaches an optimum, the sub-model's weights are frozen. The subsequent sub-models are afterwards trained iteratively taking as inputs their previous sub-model's last feature vector. 
This strategy is helpful for mitigating vanishing gradients as it allows the training of smaller parts of a bigger network. However, its main drawback is its shortsightedness as early freezing might degrade later feature representations, and thus, impede the model's expressivity.


\subsection{Joint Training}
\label{ssec:joint}

Joint training inherits from multi-objective optimisation the idea of minimising a weighted sum of the competing objective functions at play (a practice known as \textit{linear scalarisation}). For each sub-model, we attribute a loss function $\mathcal{L}_i$ as defined in \cref{eq2}. Thus, for $N$ exit stages (i.e., $N$ sub-models), the total loss can be written as a linear combination of $\mathcal{L}_i$:

\begin{equation}
\label{eq:joint_loss}
    \mathcal{L}_{tot} = \sum_{i = 0}^{N-1} \alpha_i\mathcal{L}_i\left(S, X \odot \widehat{M}_i\right)
\end{equation}

where $\alpha_i$ are weighting factors applied to each respective sub-model loss. Since we do not want to prioritise any specific exit stage, we set each $\alpha_i$ to 1.

By definition, joint training establishes information sharing between the different exit stages. The model is optimised in a multiplayer-game setting since each exit (player) strives to minimise its loss along with the best internal representations for its task. Nonetheless, the loss complexity imposed on the model leads to an accumulation of gradients from the different sub-models which may result in unstable convergence \cite{li_improved_2019}.

\section{Experimental setup}
\label{sec:setup}

Our input and target signals are 4 seconds long and pre-processed using an STFT with a window of $512$ samples ($32$ ms) and 50\% overlap, resulting in $257$ features per time frame. Baseline and joint trainings were allowed to run for up to $400$ epochs whereas layer-wise training was capped at $50$ epochs per exit stage for pre-trained models or $50 \cdot (i+1)$ epochs, where $i$ is the exit stage, for the split-layer cases. We set the learning rate to $10^{-4}$ and batch size to $512$. To prevent overfitting, we implemented early stopping with patience of $25$ epochs and decreased the learning rate by $0.9$ every $5$ epochs if there was no improvement.

\subsection{Dataset}
\label{ssec:dataset}

We trained and evaluated our models using data from the 2020 DNS Challenge \cite{reddy_interspeech_2020}. This is a synthetic dataset, composed of several other datasets (clean speech, noises, room impulse responses) that are processed so as to introduce reverberation and background noises at different SNRs and target levels, thereby allowing for arbitrary amounts of noisy-clean training examples. Similarly, the challenge provides evaluation sets --- here, we use the synthetic non-reverberant test set.

\subsection{Evaluation Metrics}
\label{ssec:eval}

For this study, we are interested in assessing the performance of our models from two perspectives: speech quality and computational efficiency. To gauge the denoised speech quality, we utilise two commonly adopted metrics, namely PESQ and DNSMOS \cite{reddy_dnsmos_2021}. For evaluating the computational efficiency of the models, we consider three measures: Floating Point Operations (FLOPs), Multiply-Accumulate Operations (MACs), and Inference Time. 

FLOPs and MACs are computed using \textit{DeepSpeed}'s built-in profiler\footnote{\url{https://github.com/microsoft/DeepSpeed}} and the TorchInfo library\footnote{\url{https://github.com/TylerYep/torchinfo}}, respectively. 
Inference time was computed on CPU for a single frame and averaged out over $1000$ samples. In all cases, we normalise the metrics to 1 second of input data, corresponding to $63$ STFT frames.

\subsection{Model Configurations}
\label{ssec:params}

\begin{table}
  \centering
  \resizebox{0.85\linewidth}{!}{\begin{tabular}{lrr}
    \toprule
    \textbf{Model class} & \textbf{Trainable params.} & \textbf{Size (FP32)} \\
    \midrule
    pretrain and baseline       & 2.78 M & 11.13 MB \\
    split\_layers  & 1.62 M & 6.48 MB  \\
    concat\_layers & 1.88 M & 7.54 MB \\
    \bottomrule
  \end{tabular}}
  \caption{Model complexity for different configurations}
  \label{tab:params}
\end{table}

We experimented with several configurations, all derived from the baseline. These comprise the baseline itself, which has been replicated and trained accordingly, as well as combinations of the techniques and adaptations mentioned earlier, and can be subdivided into the following three dimensions:

\begin{itemize}
\vspace{-0.1cm}\item
  \textbf{Exit stages}: we trained both the 6-exit and 4-exit variants introduced in \cref{ssec:arch};
\vspace{-0.1cm}\item
  \textbf{Split layers}: we trained both the straightforward early-exiting model described in \cref{ssec:arch}, starting from a checkpoint of the pre-trained baseline, as well as both split-layer variants described in \cref{ssec:splitlayers};
\vspace{-0.1cm}\item
  \textbf{Training strategies}: we employ both the layer-wise and joint training schemes described in \cref{sec:training} to determine the most suitable strategy.
\end{itemize}

To isolate the impact of each adaptation, we fix our architectural hyperparameters to the following values: for baseline and simple pre-trained early-exiting models, we adhered to the original feature sizes of $400$ units for FC$_1$, GRU$_1$, and GRU$_2$, $600$ units for FC$_2$ and FC$_3$, and $257$ units for FC$_4$ --- i.e., the number of frequency bins in the suppression mask. 
In split-layers models, the number of features in each $\Phi_{i}$ must also match the size of the suppression mask. To avoid introducing any unfair advantage into our evaluation, we picked $128$ as the size of $\Phi_{i}^*$ layers, so that the overall number of propagated features per layer is less than in the baseline.
This results in models with sizes described in \cref{tab:params}.

\section{Results}
\label{sec:results}

\begin{table*}
  \centering
  \resizebox{0.95\textwidth}{!}{\begin{tabular}{ll|rrrrrr|rrrrrr}
    \toprule
    \multicolumn{2}{c|}{} & \multicolumn{6}{c|}{\textbf{PESQ}} & \multicolumn{6}{c}{\textbf{DNSMOS}} \\
    \textbf{training} & \textbf{model $\downarrow$ \hfill exit $\rightarrow$} & \multicolumn{1}{c}{\textbf{0}} & \multicolumn{1}{c}{\textbf{1}} & \multicolumn{1}{c}{\textbf{2}} & \multicolumn{1}{c}{\textbf{3}} & \multicolumn{1}{c}{\textbf{4}} & \multicolumn{1}{c|}{\textbf{5}} & \multicolumn{1}{c}{\textbf{0}} & \multicolumn{1}{c}{\textbf{1}} & \multicolumn{1}{c}{\textbf{2}} & \multicolumn{1}{c}{\textbf{3}} & \multicolumn{1}{c}{\textbf{4}} & \multicolumn{1}{c}{\textbf{5}} \\
    \midrule
    {---} & {noisy} & \multicolumn{6}{c|}{1.58} & \multicolumn{6}{c}{3.16} \\
    {---} & {baseline} & \multicolumn{6}{c|}{2.75} & \multicolumn{6}{c}{3.88} \\
    \cmidrule(lr){1-14}
    layerwise & pretrain\_6exits       &  \textbf{1.73} &  \textbf{2.20} &  2.32 &  2.29 &  2.27 &  2.21 &   3.27 &  \textbf{3.57} &  3.68 &  3.67 &  3.66 &  3.63 \\
    layerwise & pretrain\_4exits       &  \textbf{1.73} &  2.17 & &  2.38 & &  2.38 &      3.27 &  3.56 & &  3.72 & &  3.71  \\
    layerwise & split\_layers\_6exits  &  1.71 &  2.08 &  2.36 &  2.39 &  2.37 &  2.37 & 3.26 & 3.48 & 3.65 & 3.69 & 3.68 & 3.67 \\
    layerwise & split\_layers\_4exits  &  \textbf{1.73} &  2.10 & & 2.51 & & 2.45 &     3.27 &  3.50 & & 3.75 & & 3.75 \\
    layerwise & concat\_layers\_6exits &  1.72 &  2.08 &  \textbf{2.38} &  2.43 &  \textbf{2.44} &  2.44 &   3.27 &  3.49 &  \textbf{3.72} &  3.73 &  \textbf{3.73} &  3.74 \\
    layerwise & concat\_layers\_4exits &  1.71 &  2.09 & & \textbf{2.54} & & \textbf{2.55} &     3.26 &  3.49 & & \textbf{3.77} & &  \textbf{3.78} \\
    \cmidrule(lr){1-14}
    joint     & pretrain\_6exits       &  \textbf{1.72} &  2.18 &  2.40 &  2.47 &  2.53 &  2.56 &   \textbf{3.27} &  3.45 &  3.67 &  3.73 &  3.77 &  3.80 \\
    joint     & pretrain\_4exits       &  1.70 &  \textbf{2.22} & & 2.51 & &  2.58 &   3.26 &  \textbf{3.54} & & 3.72 & & 3.78 \\
    
    joint & split\_layers\_6exits      &  1.71 &  2.13 &  2.43 &  2.50 &  2.52 &  2.52 &   3.26 &  3.45 &  3.73 &  3.76 &  3.75 &  3.76 \\
    joint & split\_layers\_4exits      &  1.69 &  2.02 & & 2.41 & &  2.42 &   3.26 &  3.42 & & 3.66 & & 3.68 \\
    joint & concat\_layers\_6exits     &  1.71 &  2.14 &  \textbf{2.44} &  2.53 &  \textbf{2.55} &  2.54 &   3.26 &  3.45 &  \textbf{3.74} &  3.77 &  \textbf{3.78} &  3.78 \\
    joint & concat\_layers\_4exits     &  1.71 &  2.13 & & \textbf{2.63} & &  \textbf{2.64} &   3.26 &  3.51 & & \textbf{3.80} & & \textbf{3.81} \\
    \bottomrule
  \end{tabular}}
  \caption{Results of all the models at different exit stages (in bold, best score for each training strategy and exit stage).}
  \label{tab:results}
\end{table*}

A full overview of the different configurations is shown in \cref{tab:results}. Here, we observe a monotonous performance increase along the exit stages until asymptotically approaching the baseline, showing that deeper layers develop more expressive representations. 
Predictably, this trend is observed across all model variants and applies to both PESQ and DNSMOS scores, with the notable advantage of smaller model size in split-layer configurations.

Split-layer designs trained in a layer-wise fashion were demonstrably effective at reducing the performance gap with the baseline, despite using fewer trainable parameters and operations than the straightforward variants. 
Indeed, the features encoded by the auxiliary layers are beneficial to the denoising tasks, especially at later exit stages (see bold figures in \cref{tab:results}) where we derive richer features.
Layer concatenation (\cref{fig:splitlayers}.b) presents the highest scores, confirming that the auxiliary pipeline benefits from mask-related features.


Overall, joint training accounts for the most impact on performance. As presented in \cref{ssec:joint}, here we allow the model to take full advantage of the freedom given by all its parameters to find the best representations for all exit stages. Here we also notice that later exit stages benefit more from this training strategy, which could be addressed by using different $\alpha_i$ in \cref{eq:joint_loss}.

\begin{figure}[t]
\begin{minipage}[b]{1.0\linewidth}
  \centering
  \centerline{\includegraphics[width=1.0\linewidth]{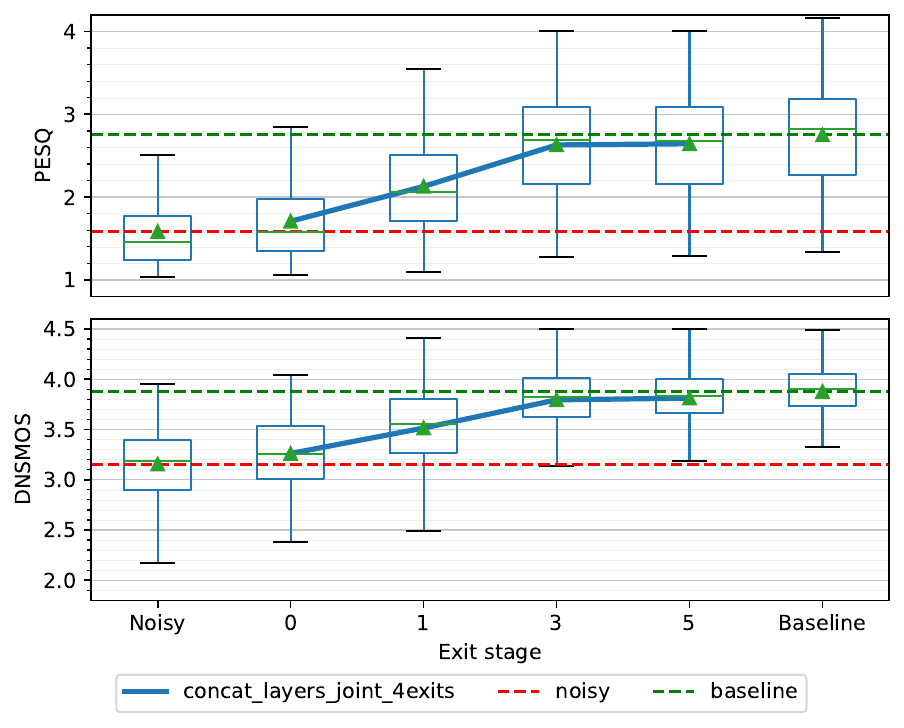}}
  \caption{Boxplot of quality metrics at different exit stages.}
  \label{fig:boxplot}
\end{minipage}
\end{figure}

In \cref{fig:boxplot}, it is interesting to observe that PESQ scores exhibit heteroskedasticity with respect to the exit stage; this can be seen as a generalisation of the baseline behaviour, which shows the largest variance, hinting that very low-quality input data are harder to recover. Bizarrely enough, the opposite phenomenon is observed when considering DNSMOS, where values are more stable around the mean.

\begin{figure}[t]
\begin{minipage}[b]{1.0\linewidth}
  \centering
  \centerline{\includegraphics[width=1.0\linewidth]{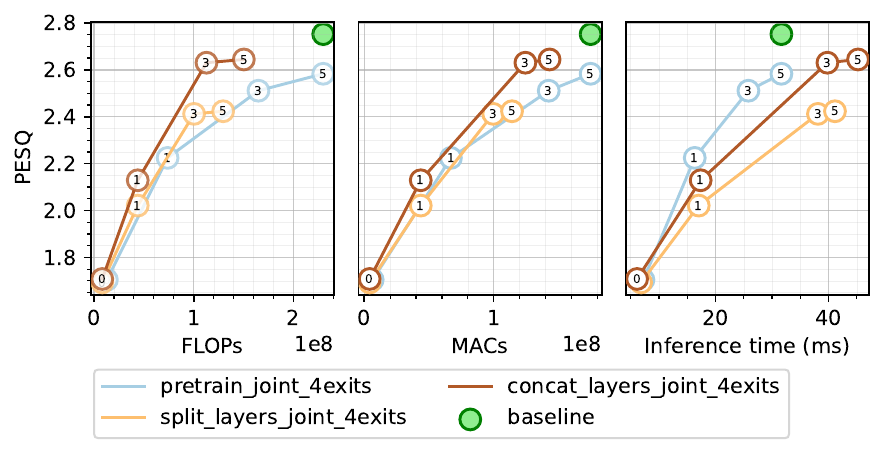}}
  \caption{Comparison of performance/efficiency trade-off at different exit stages, relative to 1 second of data.}
  \label{fig:macs}
\end{minipage}
\end{figure}


\cref{fig:macs} shows the different computational demands imposed by each layer, in terms of both operations and processing time. Expectedly, the GRU layers occupy the majority of the computational budget. 
Although splitting the layers into main and ancillary paths requires less computation than the baseline, we notice a slight increase in inference time. 
This could be caused by how Pytorch schedules computations across the layers, or by additional retrieval and copy operations.
The GRU layers also take up the majority of the inference time, due to the sequential nature of their computation. This also causes the split-layer variation to be moderately slower since they feature more GRU layers. However, exiting at a given stage $i$ will spare the computational cost of the following layers as well as that of its respective ancillary layer $\Phi_{i}^*$. Moreover, as mentioned earlier, the layers $\Phi_{i}^*$ are of small dimensions by design, thus presenting negligible overhead (see \cref{tab:params}). 

\begin{figure*}[!t]
   \centering
  \includegraphics[width=0.95\textwidth]{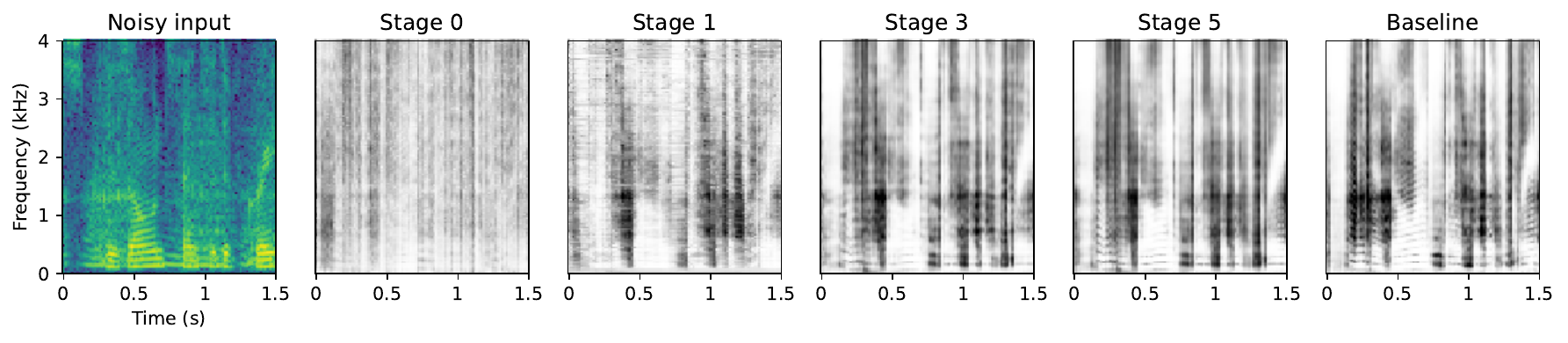}
  \caption{Comparison of example noisy input (in color), suppression masks (in grayscale) at different exits, and baseline.}
  \label{fig:spectrog}
\end{figure*}

The incremental improvements provided by each processing block can be fully appreciated in \cref{fig:spectrog}. Most noticeably, stage 0 extracts a coarse spectral envelope of the noise, while later stages refine the contour of speech features such as formats, their harmonics, and high-frequency unvoiced components. When comparing against the baseline, a small but noticeable compression in dynamic range is also observed; this could be a symptom of the conflicting objectives that joint training aims to optimise --- indeed, models trained layer-wise exhibit more contrast.

\begin{figure}[t]
\begin{minipage}[b]{1.0\linewidth}
  \centering
  \centerline{\includegraphics[width=0.95\linewidth]{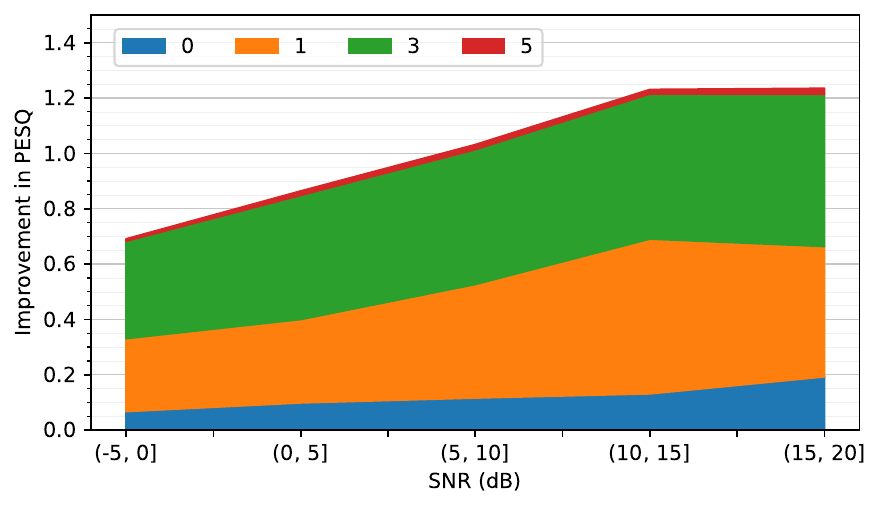}}
  \caption{Improvement in PESQ caused by each exit stage, over a range of input SNR values.}
  \label{fig:snr}
\end{minipage}
\end{figure}

Finally, in \cref{fig:snr} we provide an overview of how each model stage improves the output PESQ over different input SNR ranges. Here, we notice that the model is most effective at higher input SNRs, and that each stage's contribution to the denoising task is almost equal for each SNR bracket.


\section{Conclusion}
\label{sec:end}
In this work, we proposed an efficient and dynamic version of nsNet2, built upon early-exiting. The models presented herein provide a diverse range of performance/efficiency trade-offs, that could benefit embedded devices with computational and power constraints such as headphones or hearing aids. 
Our best-performing models can achieve 96\% and 98\% of baseline performance on PESQ and DNSMOS metrics, respectively, on the last exit stage. 
When considering the second exit stage, we are able to reach 77\% and 90\% of baseline performances with 62\% savings in multiply-accumulate operations. 

Our future work will advance the current proposed model by automatically selecting the optimal exit stage based on the properties of the current input signal as well as a non-intrusive quality assessment module.


{\small
\bibliography{refs,strings}
}
\bibliographystyle{IEEEtran}

\end{document}